# Fragmentation model for expanding cylinder

V.A. Goloveshkin[1)], N.N. Myagkov[2)1]

[1)] *Moscow State University of Instrumental Making and Information Science, Moscow, Russia*

[2)] *Institute of Applied Mechanics, Russian Academy of Sciences, Moscow, Russia*

**Abstract**
A two-dimensional energy-based model of fragmentation of rapidly expanding cylinder under plane strain conditions is proposed. The model allows us to estimate the average fragment length and the number of fragments produced by ductile fracture of the cylinder. Comparison of obtained results with published experimental data on the fragmentation of the aluminum rings and cylinders shows that they are in good agreement.

# 1.   Introduction

Two one-dimensional theories of the dynamic fragmentation of rapidly expanding metal cylinders have been actively discussed in recent years (Grady 2006). First is the statistics-based theory of Mott (Mott's theory, published in 1947, described in detail by Grady (2006). Second is the energy-based theory of Grady (Grady 2006; Kipp and Grady 1985). Mott considered the body to be rigid perfectly plastic and straining in tension under a constant flow stress. He assumed that fracture occurred instantaneously and argued that the fracture energy was not significant. Assessing fragment length he ignored fracture energy and believed that the average fragment length *s* is proportional to a size scale, which is the only

---

[1] Corresponding author. Tel.: +7-495-9461765; fax: +7-495-938 0711
   *E-mail address*: nn_myagkov@mail.ru

combination of the yield stress *Y*, density ρ and strain rate $\dot{\varepsilon}$. As a result Mott obtained

$$s = \beta_n^{-1}\left(\frac{2\pi Y}{\rho\dot{\varepsilon}^2}\frac{\sigma}{n}\right)^{n/(2n+1)}, \qquad (1)$$

where σ and *n* are the Weibull parameters, $\beta_n$ is the constant.

A key point of the energy-based fragmentation theory (Grady 2006; Kipp and Grady 1985) is an account of the energy dissipated in the fracture process. As a result of the Mott's theory generalization, Kipp and Grady (1985) obtained the following estimate for the average fragment length

$$s = \left(\frac{24\Gamma}{\rho\dot{\varepsilon}^2}\right)^{1/3}, \qquad (2)$$

where $\Gamma = Y \cdot x_{c0}/2$ is the fracture energy, $x_{c0}$ is the critical value of crack-opening displacement. The value of Γ can be expressed through the dynamic stress intensity factor $K_f = \sqrt{2E\Gamma}$ (Grady 2006), where *E* is the elastic modulus.

It is worth noting that molecular-dynamic modeling of the fragmentation of homogeneous liquid at its adiabatic expansion (Holian and Grady 1988; Ashurst and Holian 1999) also gave the two-thirds power dependence of the average fragment size on strain rate, $s \propto (\dot{\varepsilon})^{-2/3}$, regardless of the dimensionality of the system. This dependence was confirmed experimentally by Moukarzel et al. (2007).

In the present communication, a simple two-dimensional energy-based model of ductile fragmentation of rapidly expanding cylinder under plane strain conditions is proposed. The model uses a minimum number of constants characterizing the material properties of the cylinder.

## 2. Two-dimensional model of necking

A key point of our approach is a two-dimensional model of necking, in which the velocity field is similar to that occurring directly under a flat die in Prandtl's solution for problem of the indenting of a semi-infinite medium by the flat die (Prandtl 1920; Hill 1998). To model the necking we consider a two-dimensional ideal rigid-plastic "rod" with finite thickness $2h$ under plane strain conditions (Fig. 1). The rod consists of the two regions $\Sigma_1$ and $\Sigma_2$, and two equal rectangular isosceles triangles of $AOB$ and $COD$. At the initial moment the sides $AB$ and $CD$ coincide with the upper and lower boundary of the rod, respectively. Mass, power of internal forces and kinetic energy of the rod are defined per unit length in the direction perpendicular to the plane of Fig. 1.

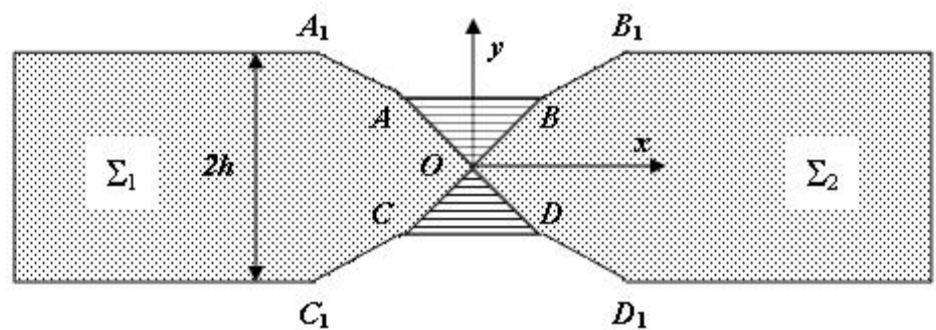

Fig. 1 Cross-section of the two-dimensional rod under plane strain conditions (the rod is infinite in the direction perpendicular to the figure plane).

The regions $\Sigma_1$ and $\Sigma_2$ (Fig. 1) move as the rigid bodies in the x-axis direction at velocities $V_1$ and $V_2$, respectively ($V_2 > V_1$). Triangle regions of $AOB$ and $COD$, which are symmetrical to each other at each time point, move as the rigid bodies in the direction of the point $O$ until they disappear. Their masses distribute in equal shares among the masses of the regions $\Sigma_1$ and $\Sigma_2$ by elongation of these regions.

We denote the velocity components of the triangle $AOB$ by $V_x$ and $V_y$. The continuity of normal velocity at the interfaces $OA$ and $OB$ gives

$$V_x = \frac{1}{2}(V_1 + V_2), \ V_y = \frac{1}{2}(V_1 - V_2). \tag{3}$$

In the framework of the model under consideration can be shown that the newly formed free surfaces $AA_1$, $BB_1$, $CC_1$ and $DD_1$ have slopes (to the $x$-axis) equal to -½ or ½ while the slopes of $OA$, $OB$, $OC$ and $OD$ are equal to -1 or 1 (Fig. 1).

We denote the absolute value of the tangential velocity discontinuities at the interfaces $OA$ and $OB$ by means of $[V]_{OA}$ and $[V]_{OB}$, respectively. Taking into account (3), we readily have

$$[V]_{OA} = \frac{\sqrt{2}}{2}\left|V_x - V_y - V_1\right| = \frac{\sqrt{2}}{2}(V_2 - V_1),$$

$$[V]_{OB} = \frac{\sqrt{2}}{2}\left|V_x + V_y - V_2\right| = \frac{\sqrt{2}}{2}(V_2 - V_1). \tag{4}$$

Owing to symmetry of the interfaces $OA$, $OB$, $OC$ and $OD$ the total power $W_{tot}$ of the internal forces on the tangential velocity discontinuities is

$$W_{tot} = 4W_{OA} = 4\int_{OA} \frac{Y}{\sqrt{3}}[V]_{OA} dL = \frac{4}{\sqrt{3}}(V_2 - V_1)\lambda Y, \tag{5}$$

where $Y$ is the yield stress and $\lambda = AB/2$. The equation for $\lambda$ has the form

$$\frac{d\lambda}{dt} = \frac{1}{2}(V_1 - V_2). \tag{6}$$

The resulting power value (5) is minimal. This statement can be proved in a more general form for deformable materials obeying the ideally plastic model. Indeed, suppose that the deformable zone is bounded by $-\lambda < y < \lambda$ and $f(y) < x < g(y)$ so that when $x > g(y)$ we have $V_x = V_2$ and $V_y = 0$, and when $x < f(y)$ we have $V_x = V_1$ and $V_y = 0$. The total power is

$$W'_{tot} = \int_{-\lambda}^{\lambda} dy \int_{f(y)-0}^{g(y)+0} Y\sqrt{\frac{2}{3}}\sqrt{\dot{\varepsilon}_x^2 + \dot{\varepsilon}_y^2 + 2\dot{\varepsilon}_{xy}^2}\, dx, \tag{7}$$

Taking into account that $\dot{\varepsilon}_x = -\dot{\varepsilon}_y$, one gets from (7)

$$W'_{tot} \geq \int_{-\lambda}^{\lambda} dy \int_{f(y)-0}^{g(y)+0} 2Y\sqrt{\frac{1}{3}}|\dot{\varepsilon}_x|dx \geq \int_{-\lambda}^{\lambda} dy \left|\int_{f(y)-0}^{g(y)+0} 2Y\sqrt{\frac{1}{3}}\frac{\partial U}{\partial x}dx\right| \geq$$

$$\geq 2Y\sqrt{\frac{1}{3}}\int_{-\lambda}^{\lambda}|U(g(y)+0) - U(f(y)-0)|dy \geq 2Y\frac{1}{\sqrt{3}}\int_{-\lambda}^{\lambda}|V_2 - V_1|dy = \frac{4}{\sqrt{3}}|V_2 - V_1|\lambda Y. \tag{8}$$

Thus $W'_{tot} \geq W_{tot}$ for the arbitrary boundaries $f(y)$ and $g(y)$.

Integrating the power (5) over time and taking into account (6) we find the fracture energy $A_p$ of the rod,

$$A_p = \int_0^{t_f} W dt = \int_h^0 W \frac{dt}{d\lambda} d\lambda = \frac{4}{\sqrt{3}} Y h^2 \qquad (9)$$

where $t_f$ is the fracture time that can be defined. Determination of the tensile force-time history expected during the necking is also of interest.

An energy balance equation has the form

$$\frac{dE}{dt} + W_{tot} = 0, \qquad (10)$$

where $E = \frac{1}{2}(m_1 V_1^2 + m_2 V_2^2)$ is the total kinetic energy, $m_1 = m_{\Sigma_1} + m_{AOB}$ and $m_2 = m_{\Sigma_2} + m_{COD}$. The masses $m_1$ and $m_2$ are constants due to the incompressibility of the rod and equal to $\rho 2 h l_{10}$ and $\rho 2 h l_{20}$, respectively, where $l_{10}$ and $l_{20}$ are the initial lengths of the left and right sides of the rod.

The equations (6) and (10) taking into account the momentum conservation law give an equation

$$\frac{d^2 \lambda}{dt^2} = \kappa^2 \lambda \qquad (11)$$

with initial conditions at $t = 0$: $\lambda = h$ and $\dfrac{d\lambda}{dt} = -\dfrac{1}{2}(V_{20} - V_{10})$, where $\kappa^2 = \dfrac{Y}{\sqrt{3}\rho h} \cdot \dfrac{l_{10} + l_{20}}{l_{10} l_{20}}$, $V_{20} = V_2(0)$ and $V_{10} = V_1(0)$. The fracture time is defined from condition at $t = t_f$: $\lambda = 0$. The tensile force (per unit length) is defined as $F = 2\lambda \dfrac{2}{\sqrt{3}} Y$. The solution of equation (11) gives

$$F = \frac{4}{\sqrt{3}} Y (h \cosh(\kappa t) - \frac{V_{20} - V_{10}}{2\kappa} \sinh(\kappa t)), \ t \in [0, t_f], \qquad (12)$$

$$t_f = \frac{1}{\kappa} \tanh^{-1}\left(\frac{2\kappa h}{V_{20} - V_{10}}\right). \qquad (13)$$

The relation (12) can be presented in the form

$$F = \frac{4}{\sqrt{3}} Yh \frac{\sinh(\kappa (t_f - t))}{\sinh(\kappa t_f)}, \ t \in [0, t_f] \qquad (14)$$

From (13) one can see that the fracture of rod occurs at condition

$$\frac{2\kappa h}{V_{20} - V_{10}} < 1. \qquad (15)$$

The solutions (9), (13) and (14) are applied below to modeling of fragmentation of rapidly expanding cylinder.

# 3. Fragmentation model for rapidly expanding cylinder

We assume that material of the cylinder is incompressible with the density ρ and its mechanical behavior obeys the ideal rigid-plastic model with yield stress *Y*. We also assume that that the wall thickness 2*h* of the cylinder is much smaller than the cylinder radius, i.e., 2*h*<<*R* (Fig. 2). The cylinder undergoes uniform circumferential stretching at a constant strain rate given by the ratio $\dot{\varepsilon}_\varphi = V_R / R$, where $V_R$ is the radial velocity.

Let us consider a sector of the cylinder (Fig. 2) with a middle line of length 2*a* and side area $\delta S = 2h$ (per unit length of the cylinder), which is supposed to detach from the cylinder forming two new surfaces. We assume that the fracture on the both ends of the sector occurs identically and estimate average length of the fragment 2*a*.

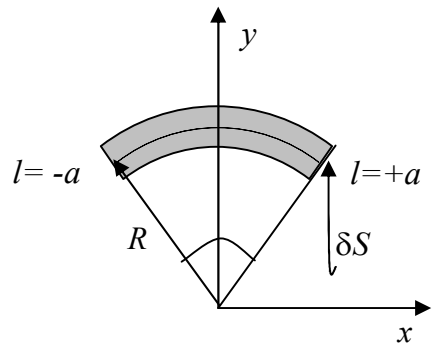

Fig. 2. Sector of the cylinder cross-section normal to the cylinder axis: *l* – distance along the middle line, $-a \leq l \leq +a$, *R* – the cylinder radius, $\delta S = 2h$ - lateral area of the fragment per unit length of the cylinder, 2*h* << *R*.

The local kinetic energy of the sector (Fig. 2), taken relatively to its center of mass is

$$T = 2h\rho\dot{\varepsilon}_\varphi^2 \frac{R^4}{a}\left(\left(\frac{a}{R}\right)^2 - \sin^2\left(\frac{a}{R}\right)\right). \tag{16}$$

For the case $a<<R$ one gets from (16)

$$T = \frac{1}{3}2h\rho\dot{\varepsilon}_\varphi^2 a^3 \tag{17}$$

The estimates show (Myagkov et al. (2009)) that the kinetic energy $T$ can be quite well approximated by the expression (17) instead of (16). Elastic energy per unit length of the cylinder stored in the fragment (before its creation) is

$$P = \frac{(\sigma_*)^2}{2E}4ah, \tag{18}$$

where $E$ is the elastic modulus and $\sigma_*$ is the critical stress at which fracture occurs. Following Glenn and Chudnovsky (1986) and Grady (1988), we assume that both the elastic and kinetic energy is spent on the creation of new fracture surfaces. Then the energy balance due to the fragmentation has the form

$$P + T = A_p. \tag{19}$$

The equation (19) takes into account that at the fragmentation of the cylinder the energy balance of the fragment must include only one fracture surface. For data taken from the experiments listed below (Table 1), the estimates made on the basis of equation (19) show that the ratio $P/T$ is small (see last column in Table 1). Therefore, instead of (19) we consider the equation

$$T = A_p. \qquad (20)$$

From (9), (17) and (20) we readily obtain the average fragment length $2a$

$$2a = \left( \frac{16\sqrt{3} \cdot Yh}{\rho \dot{\varepsilon}_\varphi^2} \right)^{1/3} \qquad (21)$$

and the average fragment number $N$

$$N = \pi R \left( \frac{\rho \dot{\varepsilon}_\varphi^2}{2\sqrt{3}\, Yh} \right)^{1/3}. \qquad (22)$$

One can see that the average fragment length (21) and the number of fragments (22) also obey the two-thirds power dependence on strain rate. However, there is a significant difference between formula (21) and formulae (1) and (2) obtained in the theories of Mott and Grady. The difference consists in the presence of the wall thickness value $h$ in the formula (21).

The tensile force is given by the formula (14) together with the fracture time

$$t_f = \frac{1}{\kappa} \tanh^{-1}\left( \frac{\kappa h}{a \dot{\varepsilon}_\varphi} \right), \qquad (23)$$

where $\kappa^2 = 2Y/(\sqrt{3}\rho h a)$. The formula (23) is obtained from (13) by the substitution of $V_{20} - V_{10} = 2a\dot{\varepsilon}_\varphi$. It is also easily seen that the condition of fracture $\dfrac{\kappa h}{a\dot{\varepsilon}_\varphi} < 1$, which follows from (15), is fulfilled identically.

Let us compare the fragment number determined by the formula (22) with that obtained in experiments by Myagkov et al. (2009), Grady and Benson (1983) and Zhang and Ravi-Chandar (2010). Type of the experiment, material used in the experiment, the experimental fragment number and the fragment number determined by the formula (22) are shown in Table 1. The material parameters are shown in Table. 2. We see that the estimates obtained by the formula (22) are in close agreement with the experimental data (last two columns in Tab. 1).

Tab. 1 Comparison with experimental data

| The experiment | Material (aluminum alloy) | $R$, mm | $2h$, mm | $\dot{\varepsilon}_\varphi$, s$^{-1}$ | The fragment number (experiment) | The fragment number (formula (22)) | $P/T$ |
|---|---|---|---|---|---|---|---|
| Ejecta destruction (Myagkov et al. 2009) | AMG6M | 11.5 | ~1.0 | $1.2 \cdot 10^5$ | 21 | 18 | 0.03 |
| Fragmentation of rings (Grady and Benson 1983) | 1100-0 | 16 | ~1.0 | $1.2 \cdot 10^4$ | 13 | 9 | 0.03 |
| | | | | $1.3 \cdot 10^4$ | 11 | 10 | 0.03 |
| | | | | $1.4 \cdot 10^4$ | 11 | 10 | 0.03 |
| Fragmentation of cylinders (Zhang and Ravi-Chandar 2010) | 6061-0 | 15.25 | 0.5 | $1.02 \cdot 10^4$ | 8 | 9 | 0.08 |
| | | | | $0.96 \cdot 10^4$ | 5 | 8 | 0.08 |
| | | | | $1.12 \cdot 10^4$ | 10 | 9 | 0.07 |
| | | | | $1.11 \cdot 10^4$ | 7-8 | 9 | 0.07 |

Tab. 2 Material data

| Material | $\rho$, g/cm$^3$ | $E$, GPa | $\sigma_*$, MPa | $Y$, MPa |
|---|---|---|---|---|
| AMG6M | 2.64 | 72 | 340 | 170 |
| Al 1100-0 | 2.71 | 68.9 | 89.6 | 34.5 |
| Al 6061-0 | 2.70 | 68.9 | 124 | 55.2 |

The data from the experiments by Myagkov et al. (2009) and Zhang and Ravi-Chandar (2010) require comments. Modeling of the experiments with ejecta (Myagkov et al. 2009) supposes that an annular layer of material flowing from the crater is fractured due to the radial velocity component. The fragment number in these experiments is fixed by number of deep holes in the low-density collector that captures the ejecta flow. We have evaluated the number of fragments in the experiments on the fragmentation of cylinders (Zhang and Ravi-Chandar 2010) using the images provided in this paper (see Fig. 14 of the paper).

## 4. Conclusions

We have proposed a simple two-dimensional energy-based model of ductile fragmentation of rapidly expanding cylinder under plane strain conditions. The mechanic behavior of the cylinder was assumed to be ideally rigid-plastic. A key point of our approach was a two-dimensional model of necking, which allowed us to obtain the value of the fracture energy (9), the tensile force-time history (14) expected during the necking and fracture criterion (15). Then, the obtained results were applied to modeling of fragmentation of rapidly expanding cylinder. As a result the average fragment length (21) and the number of fragments (22) were found. They obey the two-thirds power dependence on strain rate like the energy-based theory of Grady. However, there is a significant difference between formula (21) and Grady's formula (2). The difference consists in the presence of the wall thickness value $h$ in (21). Comparison of the fragment number determined by the formula (22) with the available experimental data, which included experiments on the fragmentation of the rapidly expanding aluminum rings (Grady and Benson (1983) and cylinders (Zhang and Ravi-Chandar 2010), and experiment on ejecta at high velocity impact (Myagkov et al. 2009), showed that they were in close agreement.

**Acknowledgements.** This research was supported by the Russian Foundation for Basic Research (project 12-01-00027) and the Program of Presidium of Russian Academy of Science (program No.25).